\documentclass[epj,referee]{svjour}                    
\usepackage{latexsym}
\usepackage{amsmath}
\begin{document}

\title{Electromagnetically Induced Vorticity Control in a Quantum Fluid Velocity Field}
\author{T. E. Raptis}
\institute{Division of Applied Technologies,\\ National Center for Science and Research,\\ Athens, Greece}
\email{rtheo@dat.demokritos.gr}

\abstract{A new method is reported by which it is possible to induce certain flux configurations 
of desired characteristics via electromagnetic means into the overall quantum probability current 
of a many-body system in the Madelung hydrodynamic picture. Some indicative applications are also 
considered with emphasis in HTC and gravitational wave research.}

\maketitle

\section{Introduction}
We know by the work of Madelung \cite{Mad1}\cite{Mad2} and Schonberg \cite{Schon} that 
the Schroedinger equation can be reduced to a pair of two "hydrodynamic" like PDEs. The pressure terms
appearing there have an intimate connection with a non-local quantum potential that appears in a similar
treatment in De Broglie - Bohm Mechanics \cite{Tsekov}. 
Madelung ideal fluid equations are directly associated with Euler hydrodynamics and vorticity theory 
while direct formal associations between ideal fluid vorticity have been less explored. 

\bigskip 
In this short report we make explicit a particular scheme that can be used in order to construct an efficient 
flux controller for the velocity field in the hydrodynamic description of quantum
many-body problems via electromagnetic means. To this aim we consider
the induced phase shift by an arbitrary vector potential and we describe a methodology for finding appropriate closed 
forms for the associated magnetic field. We also prescribe some possible applications including the case of a 
gravito-magnetic dipole induced by a modulated superfluid flow.

\section{Madelung Formalism}

One starts with the "Madelung Transform" where the complete wavefunction is decomposed in a real amplitude
and a complex phase through $\Psi(\mathbf{r},t) = R(\mathbf{r},t) \exp{\mathbf{i}S(\mathbf{r},t)}, R=|\Psi|^2$. 
Then, from the Schroedinger picture we find the system

\begin{eqnarray}
\partial_t\rho + \nabla\cdot(\rho\mathbf{u}) = 0 \\
\partial_t\mathbf{u} + \mathbf{u}\cdot\partial\mathbf{u} = -\frac{1}{\rho}\nabla\cdot\mathbf{P} - \frac{1}{m}\nabla U
\end{eqnarray}

In the above, $\rho = mR$ is the equivalent mass-charge density, $\mathbf{u}$ is the velocity field in the
quantum probability space and $\mathbf{P} = -(h^2/2m)\rho\nabla\nabla^T ln\rho$ is the corresponding non-local 
Pressure Tensor dyadic. Associated with the above picture is the quantum probability current given as

\begin{equation}
\mathbf{J} = \rho\mathbf{u} = R\nabla S
\end{equation}

The above is also expressed in the Schroedinger picture through the antisymmetric 1-tensor

\begin{equation}
\mathbf{J} = -\frac{\mathbf{i}h}{2m}\left[\Psi^{*}\nabla\Psi - \Psi\nabla\Psi^{*} \right]
\end{equation}

It is a rarely mentioned fact that both of the above experssions are identical with the so called "Monge representation"
\cite{Udriste} of arbitrary vector flows first used in hydrodynamics. Then, the vorticity vector potential can be obtained in the
so called "Clebsch variables" representation \cite{Asanov}\cite{JMars}\cite{Wagner} as

\begin{equation}
\nabla\times\mathbf{J} = \nabla R\times\nabla S = -\mathbf{i}\frac{h}{m}\nabla\Psi\times\nabla\Psi^{*} 
\end{equation}

In the above, the two alternative representations, the scalars ${R, S}$ and ${\Psi, \Psi^{*}}$ stand for the Clebsh 
"Stream" and "Flux" potentials. It should be emphasized that this representation is local and non-unique. 
Using (3) one can rewrite the first of (5) in the equivalent form

\begin{equation}
\nabla\times\mathbf{J} = \frac{1}{\rho}\nabla\rho\times\mathbf{J} = \nabla lnR\times\mathbf{J}
\end{equation}

In the case of an irrotational flow, the above shows that the current will follow the gradient of the charge-mass density.

\bigskip
Assume next that we would like to shift the total flux in a new desirable form $\mathbf{J'}$ of which the vorticity will obey a certain prescribed relation. 
Therefore, one or more external driving fields must be added. At the moment we ignore the contribution of Spin magentic moment. This is possible with the 
addition of an external non-stationary electromagnetic field having a vector potential $\mathbf{A}$ such that eq. (3) becomes

\begin{equation}
\mathbf{J} = R(\nabla S - \frac{q}{c}\mathbf{A})
\end{equation}

It is then expexted that the current flow will be driven towards the desired flow pattern. 
In order to find electric and magnetic fields compatible with the Schrodinger operator one has to solve the above conditions together with the full 
Schrodinger equation and then solve the inverse problem for the Maxwell equations to define the external currents necessary for exciting these fields. 
Hence, a self-consistent formulation of the problem can be given by any solution of the non-linear eigenvalue problem

\begin{equation}
\left[\nabla - \frac{q}{c}\mathbf{A}(\partial_i R,\partial_i S) \right]^2\Psi = \frac{2m}{h^2}(U(\mathbf{r}) - E)\Psi
\end{equation}

together with the constraints on $\mathbf{A}$ that will be prescribed in the following section.

We next examine two possible cases of such control fields, first in the case of irrotational superfluidity vortices where we show how to manipulate 
the total flux so that it would move from the ordinary cyclic vortices to a new helical type of vorticity. Then we also apply the same method to a generic
case of a hot fermion gas.

\section{Electromagnetic Current Vortification}

In the case of low temperature superfluidity \cite{},  Bose condensation obliges a global phase coherence so that the flux becomes quantized.
As a result, we get $R ~ h/m $ and the flow in (3) becomes irrotational. Using now (7) as a generic form of flux control, we want to move this flux 
into a new state satisfying a condition of the form 

\begin{equation}
\nabla\times\mathbf{J} = \lambda(\mathbf{r})\mathbf{J}
\end{equation}

In the above $\lambda(\mathbf{r})$ is an arbitrary scalar. Such a state prescribes the vorticity eigenvalue of a curl eigen-field and it is a 
characteristic of ideal helical (toroidal-poloidal) flows of classical hydrodynamics. The choice of this particular calss of flows has been made 
deliberately in order to satisfy an additional condition that causes the field to be contained inside the bulk of the flux region as explained 
further in section \textbf{4}.

\bigskip
Adapting the generic form of eq. (7) in the superflow case gives then

\begin{equation}
\mathbf{B} = \nabla\times\mathbf{A} = \lambda\nabla S 
\end{equation}

The above immediately gives the magnetic field as a function of the phase gradient and the vorticity eigenvalue. We also see that the result demands 
the magnetic field to be parallel to the flow. As superfluidity implies global coherence one may take the $S$ variable to be of 
the form $\mathbf{k}\cdot\mathbf{r}$ from which one gets $\mathbf{B} \approx \lambda(\mathbf{r})\mathbf{k}$. 

\bigskip
In the above, we did not specifically account for the contribution of the Spin degrees of freedom. These can be accounted for through an additional  
term in (7) given by $\mathbf{s} = <\Psi^{*}\mathbf{\hat{S}}\Psi>$. At the moment we restrict attention in cases where it would have no contribution 
($\nabla\times\mathbf{s} = 0$). In the case of a direct coupling with the externally applied magnetic field, it will bring about an additional 
orientational coherence of the microscopic dipole terms thus resulting in a shift of the magnetic field. This can then be accounted for by 
an additional factor in (9) as 

\begin{equation}
\mathbf{B} \approx \lambda(\mathbf{r})\mathbf{k} - \mathbf{s}
\end{equation}

Some caution is required with respect to the specific choice of $\lambda$ which has to follow the below constraints

\bigskip\textbf{1.}
The eigenvalue $\lambda$ is not allowed to be a constant as this would immediately render the magnetic field stationary as implied by (10). In the case of superfluidity we would not then be able to properly define a vector potential.

\bigskip\textbf{2.}
In order to guarantee the solenoidal character of the magnetic field ($\nabla\cdot\mathbf{B} = 0$) as well as the vanishing of the divergence 
of relation (8) we should consider the following additional constraints over $\lambda$ 

\begin{eqnarray}
\nabla(\lambda\mathbf{J})=0 \\
\nabla\lambda\cdot\mathbf{k} = 0
\end{eqnarray}

The first condition if combined with the continuity equation results in

\begin{equation}
\nabla\ln\lambda\cdot\nabla S = \partial_t( ln R)
\end{equation}

Eqs (13) and (14) are compatible only in a confined solenoidal flow that satisfies $\nabla\cdot\mathbf{J} = 0$ with a conserved total density $R$.

In the simplest possible case of a toroidal flux tube, condition(13) is guaranteed by the choice of a functional dependence of the form
$\lambda(r,z)$ as long as the wavevector $\mathbf{k}$ will be tangential in a coherent superfluid flow. A possible application of the above is discussed in section \textbf{5}. 

\bigskip
We first discuss the possibility of self-consistent solutions in the more general case of some hot fermion gas with electromagnetic control which can be discussed 
in the  same spirit of the hydrodynamic Madelung picture with a non-quantized amplitude R. Direct application of the curl into (7) and subsequent 
introduction in (8)results in the condition

\begin{equation}
\nabla R\times(\nabla S - \frac{q}{c}\mathbf{A}) - 
\frac{q}{c}R\mathbf{B} = \lambda R(\nabla S - 
\frac{q}{c}\mathbf{A})
\end{equation}

\bigskip
As the problem is quite more complicated we will hereafter examine exclusively the case of flows with constant $\lambda$ and in particular we will choose 
to set $\lambda = \lambda_0$ to simplify further examination. Relation (15) can be rewritten in the matrix form

\begin{eqnarray}
\mathbf{B} = \bar{\Gamma}\cdot\mathbf{J}_0 \\
\mathbf{J}_0 = \frac{c}{q}\nabla S - \mathbf{A} \\
\bar{\Gamma} = \frac{1}{R}[\partial R]_X - \lambda_0 \bar{I}
\end{eqnarray}

In the above, the symbol $[\partial R]_X$ stands for the matrix representation of the cross product in terms of the derivatives of R given below as

\begin{equation}
[\partial R]_X =
\begin{pmatrix}
0 & -\partial_z R & \partial_y R \\
\partial_z R & 0 & -\partial_x R \\
-\partial_y R & \partial_x R & 0
\end{pmatrix}
\end{equation}

Separating terms relating to the external fields in (15) and taking the curl via the identity 
$\nabla\times\nabla\times\mathbf{A} = - \nabla^2\mathbf{A}$ results in an inhomogeneous wave equation for the vector potential in the form

\begin{equation}
\nabla^2\mathbf{A} - \lambda_0\mathbf{A} = \nabla R\times\mathbf{A} + 
\frac{c}{q}\bar{\Gamma}\cdot\nabla S
\end{equation}

From the above we deduce that at least in the case of a single frquency mode $\omega = c\sqrt{\lambda_0}$ the additional external current sources 
should have the form

\begin{equation}
\mathbf{J}_e = \nabla R\times\mathbf{A} + 
\frac{c}{q}\bar{\Gamma}\cdot\nabla S
\end{equation}

It is now possible to insert any solution of (20) into (8) for the scheme to be self-consistent. 
Unfortunately, no known general solution of (20) exists in closed form so the problem must be tackled numerically.

\section{Field Containment in Controlled Currents}

In this section, we explain with more details the purpose behind the particular choice of the rotation eigenfields used in section \textbf{3}.
It is known for a long time, that classical Maxwell equations allow for the existence of charge - current distributions that do not radiate.
Conditions for this were first defined clearly by Goedecke\cite{Goedecke} and were later explained by Haus \cite{Haus} as a result of all 
Fourier components of such distributions being light-like thus preventing radiation emission. A macroscopic condition for radiation cancellation 
was obtained already by Devaney and Wolf \cite{Devaney} and their work was continued the last decades by Marengo and Ziolkowsky \cite{Marengo}. 
In the latter, we find a condition that should be satisfied by currents in order for them to radiate. Violation of this condition leads to the 
appearence of non-radiating elements. In explicit form this condition is written as

\begin{equation}
\nabla\times\nabla\times\mathbf{J} - k^2\mathbf{J} = 0
\end{equation} 

Direct substitution of (9) in (22) then leads to

\begin{equation}
\nabla\times(\lambda\mathbf{J}) - k^2\mathbf{J} = 0
\end{equation}

Further expansion brings the above in the form

\begin{eqnarray}
\nabla\lambda\times\mathbf{J} + \beta\mathbf{J} = 0 \\
\beta = \lambda^2 - k^2
\end{eqnarray}

Using again the matrix representation of the exterior product we may write (24) in the equivalent form

\begin{eqnarray}
\bar{G}\cdot\mathbf{J} = 0 \\
\bar{G} = [\nabla\lambda]_x + \beta\mathbf{I}
\end{eqnarray}

The above then can only satisfied if $\mathbf{J}$ is an eigenvector of the matrix form given in (19) but
with the derivatives of $\lambda$ as matrix elements. Such matrices have three distinct eigenvalues of
which one is 0 and the other two have the imaginary values 
$\pm\mathbf{i}\sqrt{1+|\nabla\lambda|^2}$. 

For the current to correspond to such an eigenvector it would have to satisfy the additional constraint $R\nabla S = \mathbf{e}_{\pm}(\partial_i\lambda)$ where $\mathbf{e}_{\pm}$ the corresponding eigenvectors. In such a case, the constraint (13) and (14) imply that one should also have $\nabla\lambda\cdot\nabla S = \nabla\lambda\cdot\mathbf{e}_{\pm}/R$ or $\mathbf{e}_{\pm}(\partial_i\lambda) = R\mathbf{k}$. 

As this is a highly nonlinear equation that restricts the choice of $\lambda$, it is expected that for most simple choices of $\lambda$ it will not be satisfied.Alternatively we may add (24) in the constraints for the choice of $\lambda$ thus guaranteeing the containement of the field components in the interior of the flux tube.

The significance of the above argument is that in this situation, the total energy of the system and the induced magnetic stresses cannot be minimised by radiative emissions thus forcing the total flux to relax into a state consistent with the externally applied field.

\section{Applications}

The method of control introduced in the previous sections could find applications in microelectronics and spintronics. It is
important to observe that in the absence of experimental data, such vortification is not known to be either beneficial or
disruptive for long range coherence. It is equally important in the case of high temperature suerconductivity to know if 
the total net vorticity of the kind explained here would be beneficial in which case there are numerous potential applications,
epsecially if it could significantly increase the transition temperatures.
Moreover, the issue of any critical spatial scales at which such vorticity could be found to contribute to long range coherence should
be treated with statistical and other methods although it is beyond the scope of the present report.  

In this section, we also examine the applicability of the method in an idealised conceptual experiment utilising superfluidity. 
Such an experiment can be described in terms of a very fast rotating confined toroidal-poloidal flux field. 
Specifically, an ideal substance to be used in such experiments with external magnetic fields is the case of a ferrofluid. 
We may then assume a dense set of appropriate cylindrical electromagnet pairs surrounding a toroidal tube. The currents in each 
pair should then be driven by a modulator circuit to approximate the field configuration found in (10) or (11) of section \textbf{3}. 
It should be noted that in actual experiments it has been found that ferrofluid condensates exhibit strong dipolar interactions. 
Yet we will not deal here with the problem of the exact contribution which requires a detailed examination of the exact interaction 
type that goes beyond the scope of the present report. We only notice that this might prove beneficial in reducing the requirements 
for the external driving field.

If this flux is reorganized accordingly to the helical flows characterizing certain solutions for an ideal Euler fluid as implied by eq. (9) 
then it might be possible to induce a certain gravitomagnetic component. In the example below we attempt a revival of an old proposal 
made by Robert Forward in his 1963 paper, "Guidelines to Antigravity" \cite{Forward}. 
There, the calculated gravitational dipole component was given as

\begin{equation}
G_D = 
-\frac{\mu_G}{4\pi}\frac{d}{dt}\left[ \mathbf{u}(\frac{r}{R})^2 \right]
\end{equation}

In the above $\mu_G = 4\pi G/c^2$ is the gravitomagnetic equivalent of the permeability, and $r, R$ the internal and external radius respectively of a confined toroidal flow. Taking the ratio $\Lambda = r/R$ constant leads to 

\begin{equation}
G_D = -\frac{G\Lambda^2}{4\pi c^2}\frac{d\mathbf{u}}{dt}
\end{equation}

The main difficulty in actually measuring such an effect stems from the value of the coefficient being of the order of $10^{-28}$. We will discuss possible ways that this could be addressed below. 

Eq (29) can be rewritten in terms of the internal variables of the ferrofluid dynamics through the direct use of (2) of which the \textit{lhs} represents the total time derivative, giving

\begin{equation}
G_D = \frac{G\Lambda^2}{4\pi mc^2}\left[ -\frac{\hbar^2}{2}\nabla\nabla^T\ln\rho + \nabla U \right]
\end{equation}

In the above we recognise the two main contributions coming from inhomogeneities in the mass flow and from the exact type of the interaction potential. In fact the first part is practically
negligible such that only the interaction potential is of importance here. We can then approximate (30) with the simpler expression

\begin{equation}
G_D \approx \frac{G\Lambda^2}{4\pi}\nabla U^* 
\end{equation}

In the above $U^* = U/mc^2$ represents the undimensioned ratio of the interaction energy to the total rest mass-energy of the circulating ferrofluid. 

Eq (31) allows an important observation. However small this ratio, it is only the spatial variations of it that contribute to the overall effect. Hence, the desired result could be enhanced by any abrupt variations of the internal distribution of the kinetic and interaction terms including the contained field quantities. Such an enhancement could also be attributed to a large number of \textit{spatial discontinuities} or flux defects that could be induced by additional means. One way to provide such a variation would then be to make the total flux inhomogeneous in the angular direction by causing a large number of "bead" like formations.

\bigskip
Despite the extreme technical difficulties of the above, it is interesting to notice that in some more modern versions of the so called Brans-Dicke or scalar-tensor theories of gravity the gravitomagnetic effect my be enhanced in the presence of EM fields by an additional mechanism.

It was recently reported by Raptis and Minotti \cite{Raptis} that in such theories there is a direct coupling between the gravitational part and the electromagnetic part such that the total energy of an intense magnetic field enters as a source for variations of the metric components via a wave equation

\begin{equation}
\nabla\Theta - \frac{1}{c^2}\partial_t^2\Theta = 
\kappa\left[ B^2 - (E/c)^2\right]
\end{equation}

Expected fluctuations of the local potential are then of the order of $c^2\nabla\Theta$. 
In \cite{Raptis} the above parameter $\kappa$ is estimated to be of the order of $10^{-4}$ so that the possibility of a combined action of both the rotating hyperfluid and the applied electromagnetic field contained inside the main circulating mass and resulting to high frequency gravitational waves, seems to be a promissing alternative for further research should these theories be verified by other experiments like the one proposed in \cite{Raptis}.

\section{Conclusions}

In this short report we examine a novel method of electromagnetically inducing a vortical structure in the velocity 
field of certain quantum fluids. As this does not appear to have ever been tested in the past, we believe that it will
open new avenues for theoretical and experimental research. An advantage of the method is that it implies certain
characteristics that are unique in a class of solutions for ideal Euler fluids. In particular, it allows control
over the geometrical and topological characteristics of such flows. Such a type of control can have important
implications in several areas including the study of high temperature superconductivity and superfluidity. An extreme
example of possible application in gravitomagnetism is also examined. Further numerical work is required for 
locating exact solutions of the proposed equations which will be given in a following report.

\end{document}